\documentclass[conference]{IEEEtran}
\IEEEoverridecommandlockouts
\usepackage{cite}
\usepackage{amsmath,amssymb,amsfonts}
\usepackage{algorithmic}
\usepackage{graphicx}
\usepackage{textcomp}
\usepackage{xcolor}
\usepackage{algorithm}
\usepackage{multirow}
\def\BibTeX{{\rm B\kern-.05em{\sc i\kern-.025em b}\kern-.08em
    T\kern-.1667em\lower.7ex\hbox{E}\kern-.125emX}}
\begin{document}

\title{Attack detection based on machine learning algorithms for different variants of Spectre attacks and different Meltdown attack implementations\\
{\footnotesize \textsuperscript{*}Note: Sub-titles are not captured in Xplore and
should not be used}
\thanks{Identify applicable funding agency here. If none, delete this.}
}

\author{
	\IEEEauthorblockN{
		Zhongkai Tong\IEEEauthorrefmark{1},
		Ziyuan Zhu\IEEEauthorrefmark{1},
		Yusha Zhang\IEEEauthorrefmark{1},
		Yuxin Liu\IEEEauthorrefmark{1},
		and Dan \IEEEauthorrefmark{1}}
	\IEEEauthorblockA{\IEEEauthorrefmark{1}Institute of Information Engineering, Chinese Academy of Sciences}
	\IEEEauthorblockA{\IEEEauthorrefmark{1}School of Cyber Security, University of Chinese Academy of Sciences, Beijing, China}
}

\maketitle

\begin{abstract}
To improve the overall performance of processors, computer architects use various performance optimization techniques in modern processors, such as speculative execution, branch prediction, and chaotic execution. Both now and in the future, these optimization techniques are critical for improving the execution speed of processor instructions. However, researchers have discovered that these techniques introduce hidden inherent security flaws, such as meltdown and ghost attacks in recent years. They exploit techniques such as chaotic execution or speculative execution combined with cache-based side-channel attacks to leak protected data. The impact of these vulnerabilities is enormous because they are prevalent in existing or future processors. However, until today, meltdown and ghost have not been effectively addressed, but instead, multiple attack variants and different attack implementations have evolved from them. This paper proposes to optimize four different hardware performance events through feature selection and use machine learning algorithms to build a real-time detection mechanism for Spectre v1,v2,v4, and different implementations of meltdown attacks, ultimately achieving an accuracy rate of over 99\%. In order to verify the practicality of the attack detection model, this paper is tested with a variety of benign programs and different implementations of Spectre attacks different from the modeling process, and the absolute accuracy also exceeds 99\%, showing that this paper can cope with different attack variants and different implementations of the same attack that may occur daily.
\end{abstract}


\section{Introduction}
In the past, hardware was considered trustworthy in information security for a long time. However, with the continuous development of cache side-channel attacks in the past decade, it poses a significant threat to the security of existing computer hardware. Researchers are constantly working on software and hardware mitigation and detection techniques to address these issues. The development of cache side-channel attacks has also made possible microarchitecture attacks such as the recently discovered Meltdown [1] and Spectre [2] attacks.

Memory isolation is a critical security mechanism provided by modern computer systems. Part of the operating system and instruction set structure prevents processes from accessing each other's memory or kernel memory and allowing different owners to run simultaneously on the same physical computing platform, such as the same cloud computing platform or desktop computing platform. Moreover, recently discovered meltdown attacks could overcome memory isolation and allow malicious processes to access memory addresses without physical and kernel memory privileges. The meltdown had a broad impact, including Intel's X86 architecture of processors, IBM POWER processors, and some ARM-based processors [3]. Malicious processes use the acceleration mechanism of out-of-order execution to cause eventual memory access requests to fail by attempting to access memory addresses that do not belong to the process. In contrast, memory access request violations are captured after the contents of the requested memory location have been loaded into the cache, and the operating system disables access to the contents of the cache by issuing the SIGSEGV signal. However, since the content is now present in the cache, a malicious process can use a cache-based side-channel attack (e.g., Flush+Reload [4]) to obtain private information in the cache.

Speculative execution is a powerful performance enhancement technique in modern processors. The technique allows a computer system to execute some tasks that may be needed in advance by guessing the future execution path. The work is completed before it is known whether it is needed to avoid possible time delays. If the guess ends up being correct, the computation results from the extra work will be committed, resulting in an additional performance gain. However, if the guessed execution path is wrong, most of the changes made by the pre-execution will be discarded, the results will be ignored, and the state of the registers will revert to their original state, but some information will be read into the cache. Spectre attacks exploit this technique to trick the processor into taking the wrong branch for the speculative execution. Furthermore, these malicious branch executions can be combined with cache-side channel attacks and eventually leak the contents of the victim's memory or registers.

To keep different computing devices safe from these hardware vulnerabilities, researchers are constantly coming up with different mitigation measures to try to solve these problems. Despite extensive efforts, the existing mitigation techniques for meltdown and ghost attacks are not perfect. Operating systems (such as Linux) map physical and kernel memory into two virtual address spaces for each running process, and malicious processes using meltdown can access any part of physical and kernel memory at any stage of their execution. Moreover, blocking or making it difficult for malicious processes to understand these mappings is crucial to mitigate meltdowns. This can be achieved using recent OS patches such as KPTI (formerly known as KAISER) [5]. However, these patches can significantly impact CPU performance of 5\% to 30\% [6].

Many software-based or hardware-based mitigation techniques have been proposed for Spectre attacks. For example, the binary-based methods oo7 [7] and Spectector [8] detect possible attacks by analyzing binaries to discover potentially vulnerable parts. Nevertheless, these methods cannot be applied to ongoing Spectre intrusions or to take action against them. Because there are many Specter attack variants, it is challenging to solve all problems with a single patch, and existing software mitigation techniques can introduce a 5\%-12\% performance loss to performance. Hardware defense methods such as DAWG [9] and InvisiSpec [10], which have been proposed in recent years, are effective against specific Spectre attacks but lack practicality and require additional hardware modifications, bringing a much more significant actual performance loss than software methods. So there is a need for a technology that has a low-performance loss, high utility, and the ability to mitigate multiple attacks and their variants.

Attack detection techniques are a more flexible form of software-based protection and are increasingly being used as the first defense against various attacks. By analyzing the specific characteristics of an attack, attack detection can provide an effective early detection mechanism, which is the theoretical basis for attack detection. Since attack detection does not require changes to the original CPU microarchitecture, it is convenient; and the performance loss associated with monitoring most process runtime specific behaviors (e.g., hardware performance counters) is shallow, so attack detection is increasingly becoming an essential method for defending against attacks.

The main contributions of this paper are as follows:
\begin{itemize}
\item In this paper, 30 different performance events regarding Spectre and Meltdown attacks are collected using hardware performance counters. The feature sets are feature preferred using random forest and principal component analysis algorithms, respectively, and the optimal feature subset of four features is obtained.
\item To the best of our knowledge, comparisons are made with other similar papers. This paper is the first unified attack detection model implemented for different variants of Spectre attacks (v1, v2, and v4) and different fusion attacks, modeled using different machine learning algorithms, and finally achieving a high accuracy rate of 99\%, indicating that the attack detection model in this paper is highly generic.
\item In contrast to other similar works, this paper is the first to collect independent benign and attack programs as validation sets for the detection mechanism in addition to the data in the test and training sets. Furthermore, it is found that the accuracy of the model in this paper can still be maintained at about 99\% on the validation set, indicating that the established attack detection model has strong practicality.
\item In addition to providing the detection accuracy and classification performance of the attack detection model, this paper further analyzes specific associations between the four selected hardware performance events and attacks.
\end{itemize}

\section{Background And Related Work}
\subsection{Spectre And Meltdown}
Meltdown attacks exploit the out-of-order execution of the microprocessor to leak secret information from the user or kernel space of the same process and other processes. Bypasses page-level privilege protection mechanisms by executing unprivileged instructions out of order. Allow even unprivileged users to read all main memory through the meltdown attack. The attack is implemented in two steps, the first step is to bypass memory isolation by unordered execution, and the second step is to obtain secret information by observing access traces in the cache using a cache-side channel attack. As attack techniques continue to evolve, multiple implementations of meltdown attacks have emerged.

Unlike Meltdown attacks, Spectre attacks do not generate any exceptions or segmentation errors but use branch prediction to bypass the isolation between user-level processes. Spectre attacks exist in various variants, mainly v1, v2, and v4 [11], and as attack techniques evolve, these have evolved into several different implementations such as Branchscope[22], Speculative buffer overflows[23], ret2spec[24], Netspectre[12], and Exspectre [13]. Spectre attacks mainly exploit branch prediction, making almost all processors vulnerable to branch prediction. Branch predictors are used to predicting conditional branch instructions, indirect branch instructions, and return stack buffers. Different variants of Spectre can exploit all three types of branch instructions. Thus, variants of Spectre attacks are still mainly implemented in two steps to achieve the attack. The first step is that the attacker misleads the CPU's branch predictor to execute unprivileged instructions speculatively; the second step is similar to meltdown, which also leaks unauthorized memory information cache side-channel attack.

\subsection{Hardware Performance Counters}
Hardware performance counters have been present in all microprocessors for over a decade. Hardware performance counters are a set of dedicated registers built into the CPU microarchitecture to collect information in the microprocessor, including cache usage information and instruction flow, and are primarily used to measure program and system performance [25]. Many researchers have used HPC for security purposes in recent years, such as malware detection, cache side-channel attack detection. Because hardware performance counters can collect multiple performance events, different program characteristics can be found during process execution. So using this principle can reflect the process security status, and it was found in previous studies that performance counters introduce very little loss, making it a near-perfect attack detection tool. The existing hardware performance counter tools are PAPI, Intel PMU, and Perf, and this paper uses Perf, a performance counter for Linux systems.

\subsection{Machine Learning Algorithms}
In information security, machine learning is applied to various areas, such as intrusion detection and traffic anomalies. In this paper, five different machine learning algorithms: LDA[14], KNN[15], logistic regression[34], SVM[35], and Adaboost[36] are selected to construct detection models for Spectre and Meltdown attacks. Machine learning algorithms with different algorithmic principles and computational complexity are modeled separately to verify the feasibility of building attack detection models with information from limited hardware performance counters. Finally, the most suitable algorithm is selected to build the attack detection models in this paper.

\subsection{Related Work}
Because of the attrition associated with some hardware and software attack defenses and their limitations, more researchers are devoting themselves to studying attack detection techniques. This section summarizes the recent developments regarding attack detection for Spectre and Meltdown attacks and discusses their advantages and disadvantages. In related work in 2018, four different hardware performance events (LLC\_references, LLC\_miss, branches, and branch\_mispredictions) were collected at a sampling rate of 100ms in paper [16] build an attack detection model against Spectre V1 and finally achieved 99.98\% accuracy. However, the sample size used in the paper is small, with a total sample size of only 2400, and the results are subject to change. In another paper [17], three hardware events (LLC misses, LLC accesses, and the total number of instructions) were collected at a sampling rate of 100ms to build an attack detection model for both v1 and v2 variants of Spectre. The attack detection model achieved an accuracy of 99.23\%. However, the positive and negative sample sizes are incredibly unbalanced, and the total number is negligible and unconvincing (317 positive samples and 1247 negative samples). In more recent work, [18] targeted the v1 and v2 variants of the Spectre attack and the meltdown attack, and the paper collected five and four different hardware performance events at a sampling rate of 100 ms to model attack detection for each of the three attacks. The main feature in the attack against Spectre is branching instructions, while the main feature against the meltdown attack is Total page faults. The absolute accuracy of the highest attack detection is obtained as 99.97\% (v1),99.98\% (v2),99.97\% (meltdown), respectively. Although excellent classification results are achieved, building different attack detection models for different attacks increases the performance loss of attack detection and lacks practicality. In [19], the paper performed attack detection for v1 and v2 of Spectre by collecting operators and operands, modeled using the LSTM algorithm, and finally obtained an accuracy of 97.2\%, but the possible false alarm rate is high, which may reach up to 12\%. The paper [20] collects eight different hardware events for v1 and v2 variants of Spectre and eventually obtains an accuracy of 98.7\%. However, many processors can only provide interfaces to collect four hardware performance events simultaneously, so it is challenging to apply it to real-time attack detection. In a recent paper on attack detection for the meltdown, the attack is detected by analyzing the broken errors caused by the meltdown attack. Low overhead and high accuracy attack detection against meltdown were achieved, but the paper does not verify whether different implementations against meltdown are still valid [21]. In the work of paper [33], by analyzing different hardware performance events, different hardware performance events are collected as feature sets through 100ms sampling interval to build attack detection models for Spectre and Meltdown, respectively, and finally achieve reasonable accuracy rates. However, the performance loss brought by the attack detection models established in the paper is too high, which may reach up to 8\%.

Moreover, the different hardware performance events used to build the models are not conducive to extending them to real-time attack detection. The attack detection against Spectre also encountered challenges in the paper [26] when four hardware performance events (using the same features as in the paper [16]) were collected using a sampling rate of 100ms to build an attack detection model, the purpose of evading attack detection could be achieved by reducing the branch prediction errors occurring within a unit event. After experiments, it is found that the accuracy of attack detection can be reduced to about 70\% while achieving a high attack success rate. To sum up, it shows that similar work in recent years has been challenging to meet the needs of attack detection, and there is an urgent need for an anti-evasive attack detection technology for multiple attacks.

\section{Model Implementation}
This section describes the process of constructing an attack detection model for Spectre v1, v2, v4, and different implementations of meltdown attacks in this paper. In this paper, the hardware performance events obtained from benign and attack programs are used as the data source for modeling. The data are randomly divided into two parts according to the ratio of 8:2. 80\% of the data is used as the training set for building the detection model, and 20\% of the data is used as the test set. A tenfold cross-validation method is used in the paper to obtain the accuracy of the detection model and try to avoid the overfitting problem brought by the modeling process. Unlike previous work, this paper adds a validation set in addition to the training and test sets. The validation set comprises hardware performance events collected by benign and attack programs independent of the modeled data sources. In this paper, the validation set is used to validate the model to illustrate the usefulness of the established attack detection model and make the obtained results more objective and accurate.

The steps of the attack detection model constructed in this paper are as follows.
\begin{itemize}
\item Datd extraction: For various benign and attack processes, hardware performance events of different processes are collected at 1ms sampling intervals as data sources.
\item Feature selection: For the data collected in the first step, feature preference is performed, unnecessary features are removed, and the data source is updated.
\item Data partitioning: The data source obtained in the first step is randomly divided into two parts of 8:2. 80\% of them are used as the training set and 20\% as the test set.
\item Model training: The training data is modeled using different machine learning algorithms to obtain the sub-accuracy of the training set of the detection model.
\item Model testing: Repeat step 3 to obtain the optimal parameters of the model by adjusting the main parameters of different algorithms. Finally, the data from the test set are used to test the model built from the training set to obtain the attack detection accuracy.
\item Classification results: We use the parameters obtained in step 5 to model and perform 10-fold cross-validation and calculate the average accuracy as the absolute accuracy of the classification model.
\item Model validation: Hardware performance counter information is collected for different benign processes and different attack implementations to form the validation set. Moreover, use the validation set to verify the utility of the attack detection model built above.
\end{itemize}

Figure 1 shows the overall experimental flowchart of the attack detection model built in this paper for different Spectre variant attacks and different Meltdown attack implementations.
\begin{figure}[htbp]
\centering
\includegraphics[width=3.5in,height=4.0in]{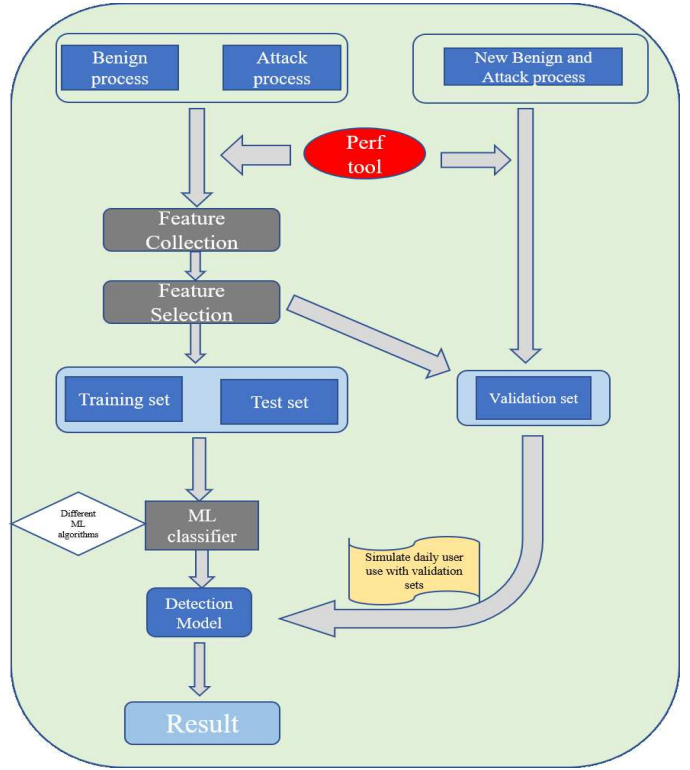}
\caption{Sample distribution of page\_fault by different processes} \label{fig1}
\end{figure}
\section{Experiment}
\subsection{Experimental Setup}
The hardware platform used for the experiments in this paper is Intel Core I5-7200U, with a processing frequency of 2.5GHz equipped with a 3Mb cache and 8GB of RAM. The operating system chosen for the experiments is Ubuntu 16.04. In this paper, the different hardware performance events are collected in real-time using the function (perf\_event\_open) provided by perf.

\subsection{Data Set}
To train a supervised learning model with utility, we created a dataset consisting of hardware performance counter data for various benign and attack processes during operation. The data information is collected for each process rather than for the overall readings of the entire CPU. The purpose of doing so is to make attack detection accurate to a single process, which is more practical than extending it to the whole system. In this paper, the attack detection system can determine whether a process is benign by performing attack prediction on the specific process running and notify the suspicious process to the user or terminate the malicious process directly.

Previous works have given certain features directly and collected limited information on hardware performance events to build attack detection models, lacking the process of objective feature preference, so this paper collects 25 hardware performance events related to different attacks as the original set of features. Table 1 shows all the features collected in this paper.
\begin{table*}[]
\caption{Collection of all hardware performance events}
\begin{tabular}{|l|l|l|l|}
\hline
Perf &
  PERF\_TYPR\_HARDWARE &
  PERF\_TYPE\_SOFTWARE &
  PERF\_TYPE\_HW\_CACHE \\ \hline
Event &
  \begin{tabular}[c]{@{}l@{}}1.PERF\_COUNT\_HW\_CAC HE\_REFERENCES\\ 2.PERF\_COUNT\_HW\_CAC HE\_MISSES\\ 3.PERF\_COUNT\_HW\_CPU\_CYCLES\\ 4.PERF\_COUNT\_HW\_INSTRUCTIONS\\ 5.PERF\_COUNT\_HW\_BUS \_CYCLES\\ 6.PERF\_COUNT\_HW\_REF\_CPU\_CYCLES\\ 7.PERF\_COUNT\_HW\_BRANCH\_MISSES\\ 8.PERF\_COUNT\_HW\_CACHE\_MISSES\end{tabular} &
  \begin{tabular}[c]{@{}l@{}}9.PERF\_COUNT\_SW\_CPU\_CLOCK\\ 10.PERF\_COUNT\_SW\_TASK\_CLOCK\\ 11.PERF\_COUNT\_SW\_PAGE\_FAULTS\\ 12.PERF\_COUNT\_SW\_CONTEXT\_SWITCHES\end{tabular} &
  \begin{tabular}[c]{@{}l@{}}13.PERF\_COUNT\_HW\_CACHE\_L1D\_READ\\ 14.PERF\_COUNT\_HW\_CACHE\_L1D\_WRITE\\ 15.PERF\_COUNT\_HW\_CACHE\_L1D\_ACCESS\\ 16.PERF\_COUNT\_HW\_CACHE\_\_L1D\_MISS\\ 17.PERF\_COUNT\_HW\_CACHE\_L1I\_MISS\\ 18.PERF\_COUNT\_HW\_CACHE\_LL\_READ\\ 19.PERF\_COUNT\_HW\_CACHE\_LL\_WRITE\\ 20.PERF\_COUNT\_HW\_CACHE\_LL\_ACCESS\\ 21.PERF\_COUNT\_HW\_CACHE\_LL\_MISS\\ 22.PERF\_COUNT\_HW\_CACHE\_DTLB\_READ\\ 23.PERF\_COUNT\_HW\_CACHE\_DTLB\_WRITE\\ 24.PERF\_COUNT\_HW\_CACHE\_DTLB\_ACCESS\\ 25.PERF\_COUNT\_HW\_CACHE\_DTLB\_MISS\\ 26.PERF\_COUNT\_HW\_CAC HE\_ITLB\_ACCESS\\ 27.PERF\_COUNT\_HW\_CAC HE\_ITLB\_MISS\\ 28.PERF\_COUNT\_HW\_CACHE\_BPU\_READ\\ 29.PERF\_COUNT\_HW\_CACHE\_BPU\_ACCESS\\ 30.PERF\_COUNT\_HW\_CACHE\_\_BPU\_MISS\end{tabular} \\ \hline
\end{tabular}
\end{table*}

In their previous work [26], they experimentally found that building an attack detection model using hardware event data collected at a sampling rate of 100ms was able to reduce the attack detection accuracy to 70\% and achieve a high attack success rate at the same time by using different ways of evasion detection. In this paper, a sampling rate of 1ms is used to collect different hardware performance events to avoid attack evasion, and the high sampling rate of hardware counters may bring too much loss of high performance []. In this paper, data from the following 12 different scenarios are collected to generate a large dataset with a balance of positive and negative samples.
\begin{itemize}
\item Stress\_C: The process repeatedly keeps calculating the square root of a random number[27].
\item Stress\_M: The process keeps calling the memory allocation function malloc and the memory release function free[27].
\item Stress\_I: The process calls sync() repeatedly. Sync () is used to write the contents of memory to the hard disk[27].
\item Firefox: Use firefox to browse the web randomly and perform random operations such as searching.
\item Video: Use a video player to play videos.
\item Meltdown: This paper uses the meltdown attack, including meltdown, meltdown\_nonull, and meltdown\_fast, three different ways to implement the attack [28].
\item Spectre: This paper uses specter attacks, including spectre V1(1), V2, and V4, three different attacks [29].
\end{itemize}

The above dataset constitutes all the data for the training and testing sets required for the attack detection modeling in this paper, with an overall sample size of 152156. The amount of data of Stress\_C, Stress\_M, Stress\_I are 15000, the amount of data of Firefox and Video are 20000, the amount of data of v1 is 11024, the amount of data of v2 is 10305, the amount of data of M is 10664, the amount of data of MF is 10598, the amount of data of MN is 10564, the amount of data of v4 is 14001.
\begin{itemize}
\item Stress:  Includes both Stress\_C and Stress\_M parts. These two parts are different from the data collected in the test set and training set.
\item MiBench: Uses various test sets from the MiBench benchmark collection, including: Basic, Bit, Cjpeg, CRC, Dijkstra, FFT, Gsm, Patrica, Qsort, Susan, and Typeset[30].
\item Spectre: Unlike the test set and training set data, using another code implementation of spectre V1(2) attack [2].
\end{itemize}
This paper incorporates a validation set and a training set, and a test set, unlike previous work. The utility of the attack detection model is validated with a separate validation set, and the different scenarios included in the validation set are shown above. The total sample size of the validation set is 34838. Furthermore, the labels of all data in this paper are 0 for all benign programs, 1 for Spectre v1, 2 for Spectre v2, 3 for different implementations of meltdown, and 4 for Spectre v4, respectively.
\subsection{Feature Selection}
Since multiple features are acquired, and in the existing system, only 4-8 sampling interfaces are generally provided, in order to be able to apply the attack detection system to real-time monitoring, and considering the performance loss caused by multi-event sampling, this paper makes the preferential selection of the acquired features. It selects no more than four features to build the attack detection model.

In previous similar work, most would assume that Spectre attacks can be detected by analyzing information about branch predictors [16 18 20 26 32], and meltdown attacks can be detected by analyzing information about page fault [18 33]. However, this paper finds that this is not the case after analyzing the data obtained in different scenarios.
\begin{figure}[htbp]
\centering
\includegraphics[width=2.8in,height=2.3in]{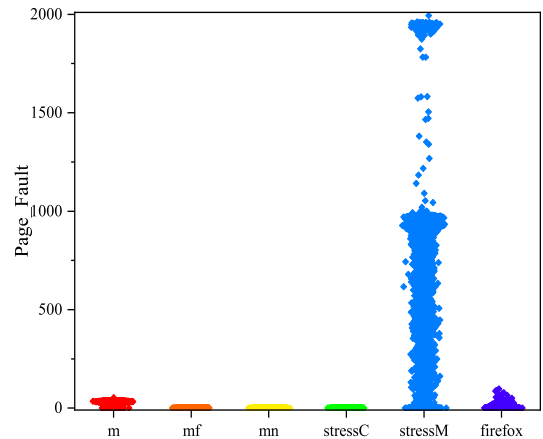}
\caption{Sample distribution of page\_fault by different processes} \label{fig1}
\end{figure}

It is evident from figure 2 that firefox and Stress\_M are the same as the general meltdown attack; both generate a large number of page\_fault. Furthermore, for different attack implementation methods, the performance of the feature page\_fault is also different. Meltdown\_nonull and Meltdown\_fast in the attack implementation process do not generate page table errors, the same as ordinary benign procedures. So page\_fault is not suitable as a feature to discover meltdown attacks.

This paper uses two features, BPU\_ACCESS and BPU\_MISS, to analyze whether the Spectre attack can be detected by information related to the branch predictor.
\begin{figure}[htbp]
\centering
\includegraphics[width=3.2in,height=2.3in]{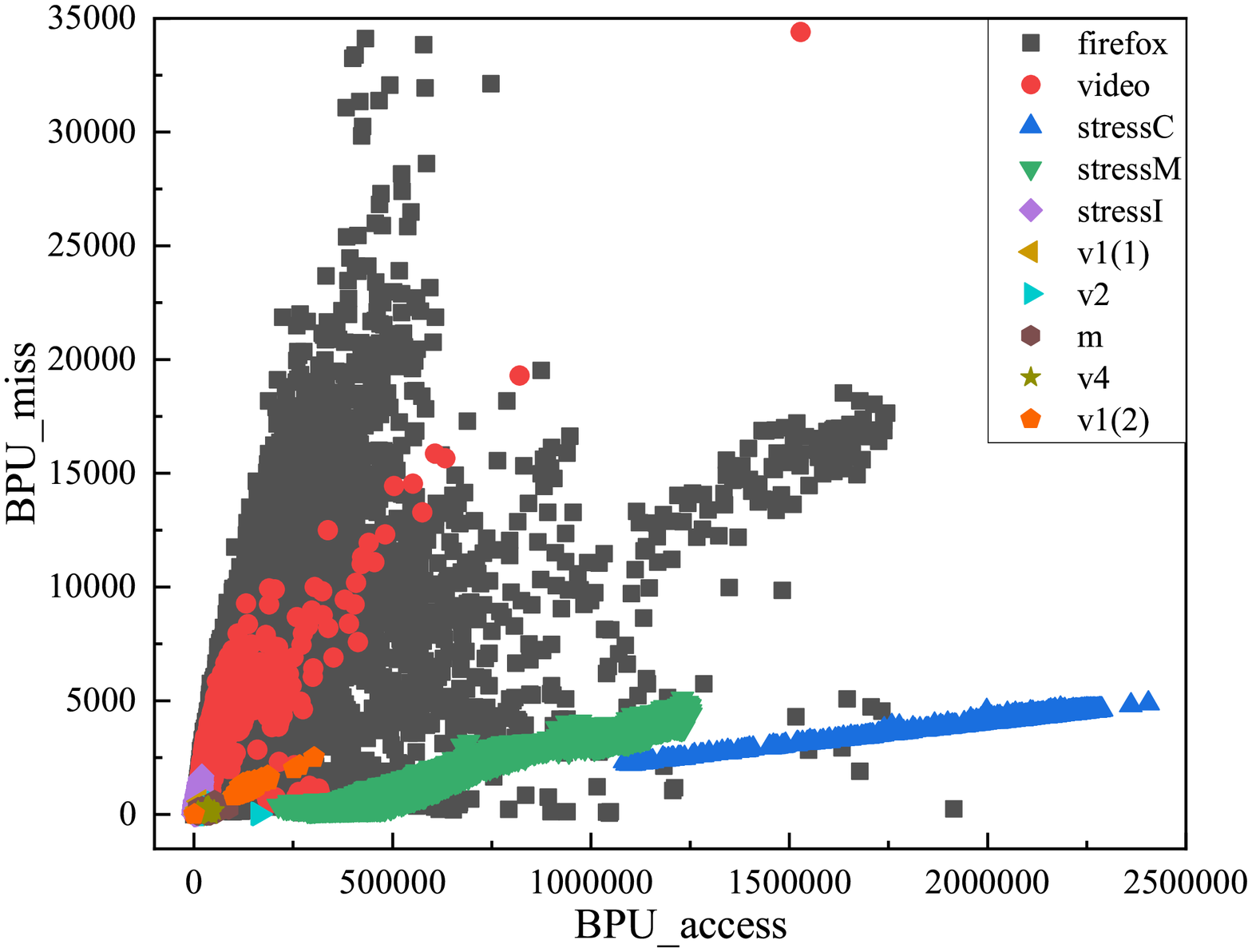}
\includegraphics[width=3.2in,height=2.3in]{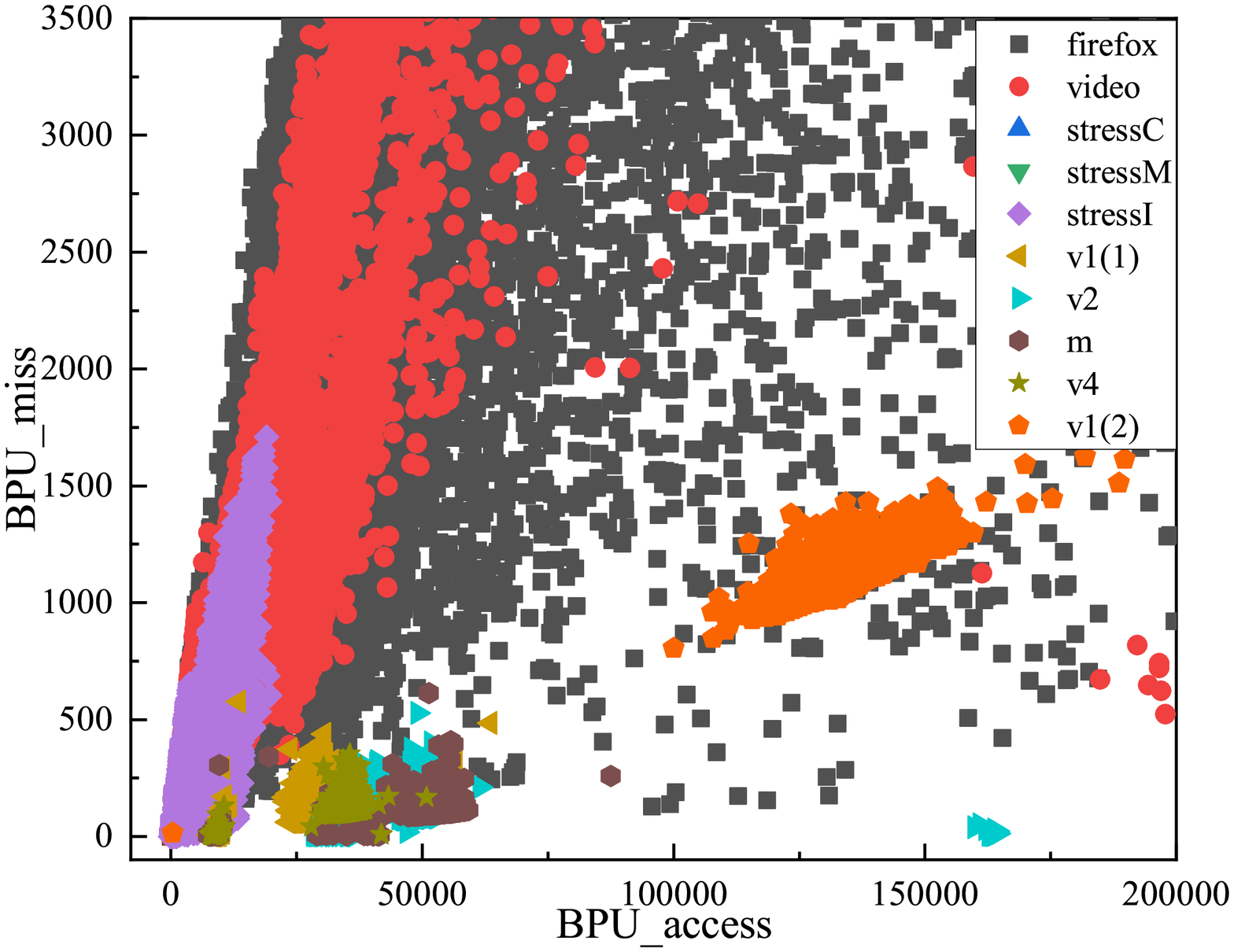}
\caption{Sample distribution of BPU features collected by different processes}\label{fig2}
\end{figure}

Figure 3 shows the sample distribution of benign and attack programs for BPU\_ACCESS and BPU\_MISS. It is obvious to find that for firefox and video, which are highly occupied programs, their distribution is extensive, and there is a significant overlap with the sample distribution of different Spectre attacks. So the information about BPU is challenging to distinguish the difference between benign and attack programs.

In this paper, four of the remaining features were selected as the feature set using the random forest algorithm for the final selection. Furthermore, the basic idea of random forest filtering features is to observe how much the accuracy is affected after modeling with and without features. To make the results of feature preferences more intuitive, this paper compares them by numerically parameterizing the difference between the accuracy before and after adding features and adding features in a ratio operation. Table 2 shows the influence size of the four characteristics with the highest importance.
\begin{table}[htbp]
\caption{The impact factors of the four features with the highest correlations}
\begin{center}
\begin{tabular}{|c|c|}
\hline
Feature     & Socre  \\ \hline
LL\_ACCESS  & 0.2467 \\ \hline
L1D\_WRITE  & 0.2456 \\ \hline
DTLB\_WRITE & 0.1053 \\ \hline
DTLB\_READ  & 0.0350 \\ \hline
\end{tabular}
\end{center}
\end{table}

In summary, the feature sets used in this paper are LL\_ACCESS, L1D\_WRITE, DTLB\_WRITE, and DTLB\_READ.
\subsection{Evaluation Metrics}
The following describes the four evaluation metrics used in this paper for the attack detection model. Where TP is the number of positive classes predicted as positive, FN is the number of positive classes predicted as negative, FP is the number of negative classes predicted as positive, and TN is the number of negative classes predicted as negative.
\begin{gather}
Auccracy=\frac{TP+TN}{TP+FP+TN+FN}Precision=\frac{TP}{TP+FP}\nonumber\\
F1\_scores=\frac{2}{{\frac{1}{Precision}+\frac{1}{Recall}}}Recall=\frac{TP}{TP+FN}\nonumber
\end{gather}
\subsection{Result}
This paper uses four hardware events as the feature set of the model for attack detection and tries different machine learning algorithms to build a unified detection model for different variants of Spectre and Meltdown attacks. Table 3 shows the accuracy of attempting to build attack detection models using different machine learning algorithms.
\begin{table}[htbp]
\caption{Accuracy obtained by modeling with different machine learning algorithms}
\begin{center}
\begin{tabular}{|l|l|}
\hline
ML algorithm & Accuracy \\ \hline
LDA          &  72.89\%        \\ \hline
LR           &  81.90\%        \\ \hline
KNN          &  63.95\%        \\ \hline
SVM          &  91.18\%        \\ \hline
Adaboost     &  99.88\%        \\ \hline
\end{tabular}
\end{center}
\end{table}

From the contents of Table 3, it is found that Adaboost obtains the best results. Therefore, the Adaboost algorithm is finally chosen in this paper to analyze further the other classification performance of the attack detection model. Table 4 shows the final obfuscation matrix obtained using the test set for the attack detection model built based on the Adaboost algorithm in this paper.
\begin{table}[htbp]
\caption{Confusion matrix obtained from the test set}
\begin{center}
\begin{tabular}{|l|l|l|l|l|l|}
\hline
Confusion Matrix & Benign & V1   & V2   & M    & V4   \\ \hline
Benign           & 17019  & 2    & 1    & 3    & 0    \\ \hline
V1               & 0      & 2243 & 0    & 4    & 1    \\ \hline
V2               & 0      & 0    & 2046 & 1    & 0    \\ \hline
M(M,MF,MN)       & 0      & 6    & 0    & 6280 & 18   \\ \hline
V4               & 0      & 1    & 0    & 5    & 2802 \\ \hline
\end{tabular}
\end{center}
\end{table}

Because the smaller the number of collected hardware performance events, the less performance loss, this paper tries to remove the least noteworthy feature among the four features, DTLB\_READ, and use three hardware performance events to build the attack detection model, and finally, the accuracy of the model decreases from 99\% to 96\%. Because the attack is destructive and has a broad impact, to maintain the high accuracy of the attack detection, this paper finally still selects four hardware performance events as the set of features.
\begin{table}[htbp]
\caption{Classification performance obtained by modeling the Adaboost algorithm}
\begin{center}
\begin{tabular}{|l|l|}
\hline
Classification Performance & Accuracy \\ \hline
Precision         & 1.00         \\ \hline
Recall            &  1.00        \\ \hline
F1\_scores        & 1.00         \\ \hline
\end{tabular}
\end{center}
\end{table}

The performance of the attack detection model from Table 5 shows that the Adaboost-based attack detection model proposed in this paper achieves excellent detection results for different Spectre attack variants and meltdown attacks, proving that malicious processes in the system can be accurately detected with limited information. In order to further validate the situation in actual use and simulate whether users can accurately detect potential attacks during daily use, this paper validates the attack detection model using different benign procedures and different attack implementations, and the different attack implementations are chosen further to test the robustness of the attack detection model. Table 5 shows the confusion matrix obtained by using the validation set to verify the attack detection model, and the accuracy of the attack detection model in this paper is 99.76\% on the validation set.
\begin{table}[htbp]
\caption{Confusion matrix obtained from the validation set}
\begin{center}
\begin{tabular}{|l|l|l|l|l|l|}
\hline
Confusion Matrix & benign & V1    & V2 & M & V4 \\ \hline
Benign           & 11047  & 12    & 30 & 0 & 0  \\ \hline
V1               & 40     & 23709 & 0  & 0 & 0  \\ \hline
V2               & 0      & 0     & 0  & 0 & 0  \\ \hline
M(M,MF,MN)       & 0      & 0     & 0  & 0 & 0  \\ \hline
V4               & 0      & 0     & 0  & 0 & 0  \\ \hline
\end{tabular}
\end{center}
\end{table}

Table 6 shows the comparison between the results of this paper and similar work. It is found that this paper builds a unified attack detection model for multiple attacks by using limited hardware performance events as features and obtains excellent attack detection accuracy.
\begin{table*}[htbp]
\caption{Comparison with the results of similar work.}
\begin{center}
\begin{tabular}{|l|l|l|c|c|}
\hline
Paper &
  Attack Types &
  Accuracy &
  \multicolumn{1}{l|}{Feature number.} &
  \multicolumn{1}{l|}{Number of samples} \\ \hline
Paper[16]                   & Spectre V1                                                                 & 99.98\% & 4                  & 2400                    \\ \hline
Paper[20]                  & \begin{tabular}[c]{@{}l@{}}Spectre V1\\ Spectre V2\\ Meltdown\end{tabular} & 98.7\%  & 8                  & NA                      \\ \hline
Paper[19]                 & \begin{tabular}[c]{@{}l@{}}Spectre V1\\ Spectre V2\end{tabular}            & 97.2\%  & NA                 & NA                      \\ \hline
Paper[21]                 & Meltdown                                                                   & 99\%    & 4                  & NA                      \\ \hline
\multirow{3}{*}{Paper[18]}    & Spectre V1                                                                 & 99.80\% & \multirow{2}{*}{5} & \multirow{3}{*}{100000} \\ \cline{2-3}
                         & Spectre V2                                                                 & 99.98\% &                    &                         \\ \cline{2-4}
                         & Meltdown                                                                   & 99.97\% & 4                  &                         \\ \hline
Paper[17]                & \begin{tabular}[c]{@{}l@{}}Spectre V1\\ Spectre V2\end{tabular}            & 99.23\% & 5                  & 15635                   \\ \hline
\multirow{2}{*}{Paper[32]} & Spectre V1                                                                 & 99.97\% & 6                  & \multirow{2}{*}{NA}     \\ \cline{2-4}
                         & Meltdown                                                                   & 99.97\% & 5                  &                         \\ \hline
Our Result &
  \begin{tabular}[c]{@{}l@{}}Spectre V1\\ Spectre V2\\ Spectre V4\\ Meltdown\\ Meltdown\_Fast\\ Meltdown\_Nonull\end{tabular} &
  99.88\% &
  4 &
  186994 \\ \hline
\end{tabular}
\end{center}
\end{table*}

The ROC plot in Figure 4 shows the performance of the attack detection model built in this paper based on the Adaboost algorithm to cope with different variants of Spectre's attack and Meltdown attack. The ROC curve shows the rate of actual positive samples being misclassified as false-positive samples in the attack detection model. The value of AUC in the figure is the area under different ROC curves, which is usually used as a criterion to evaluate the model's performance. Because ROC curves sometimes do not directly reflect the performance of a model, while AUC as a numerical value allows the performance to be quantified and expressed more intuitively. It is evident in Figure 5 that the attack detection model built in this paper shows excellent detection performance regardless of the type of attacks targeted.
\begin{figure}[htbp]
\centering
\includegraphics[width=3.6in,height=2.8in]{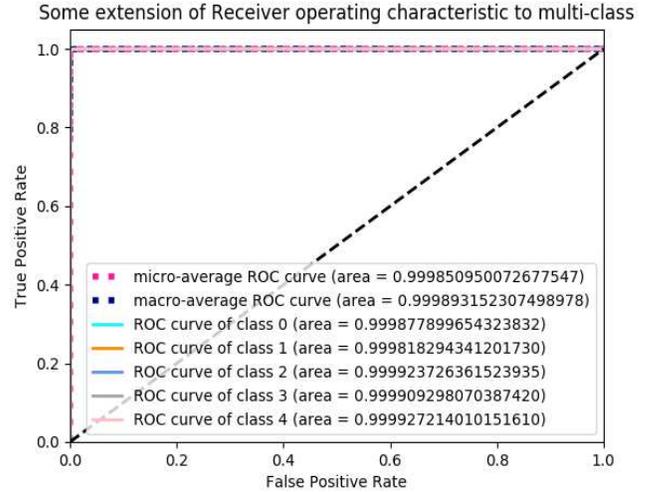}
\caption{ROC plot of the attack detection model built by Adaboost algorithm} \label{fig1}
\end{figure}

\section{Conclusion And Discussion}

\begin{figure*}[htbp]
\begin{minipage}[t]{0.5\linewidth}
\centering
\includegraphics[width=3.5in]{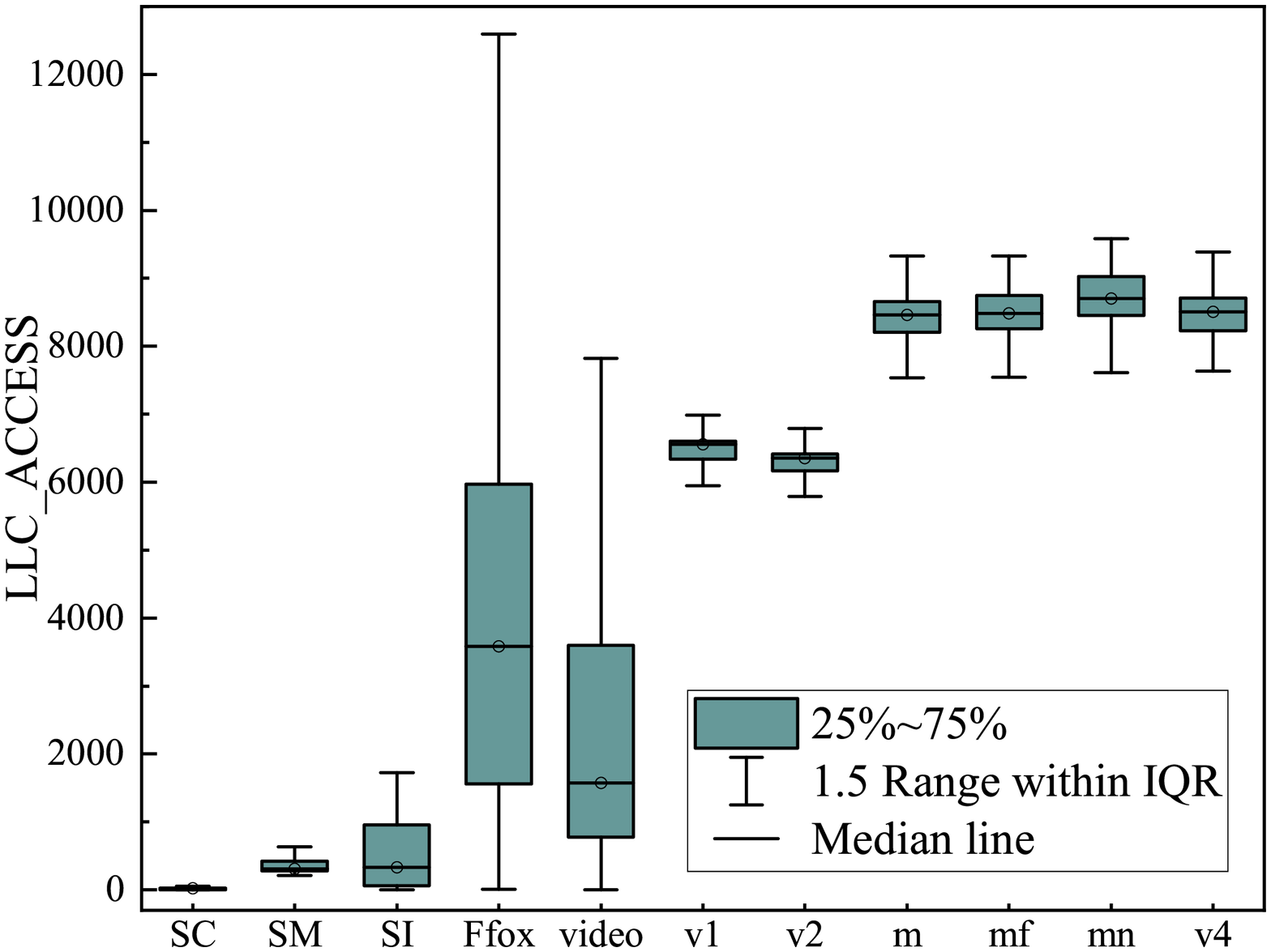}
\centerline{$LLC\_ACCESS$}
\end{minipage}
\begin{minipage}[t]{0.5\linewidth}
\centering
\includegraphics[width=3.5in]{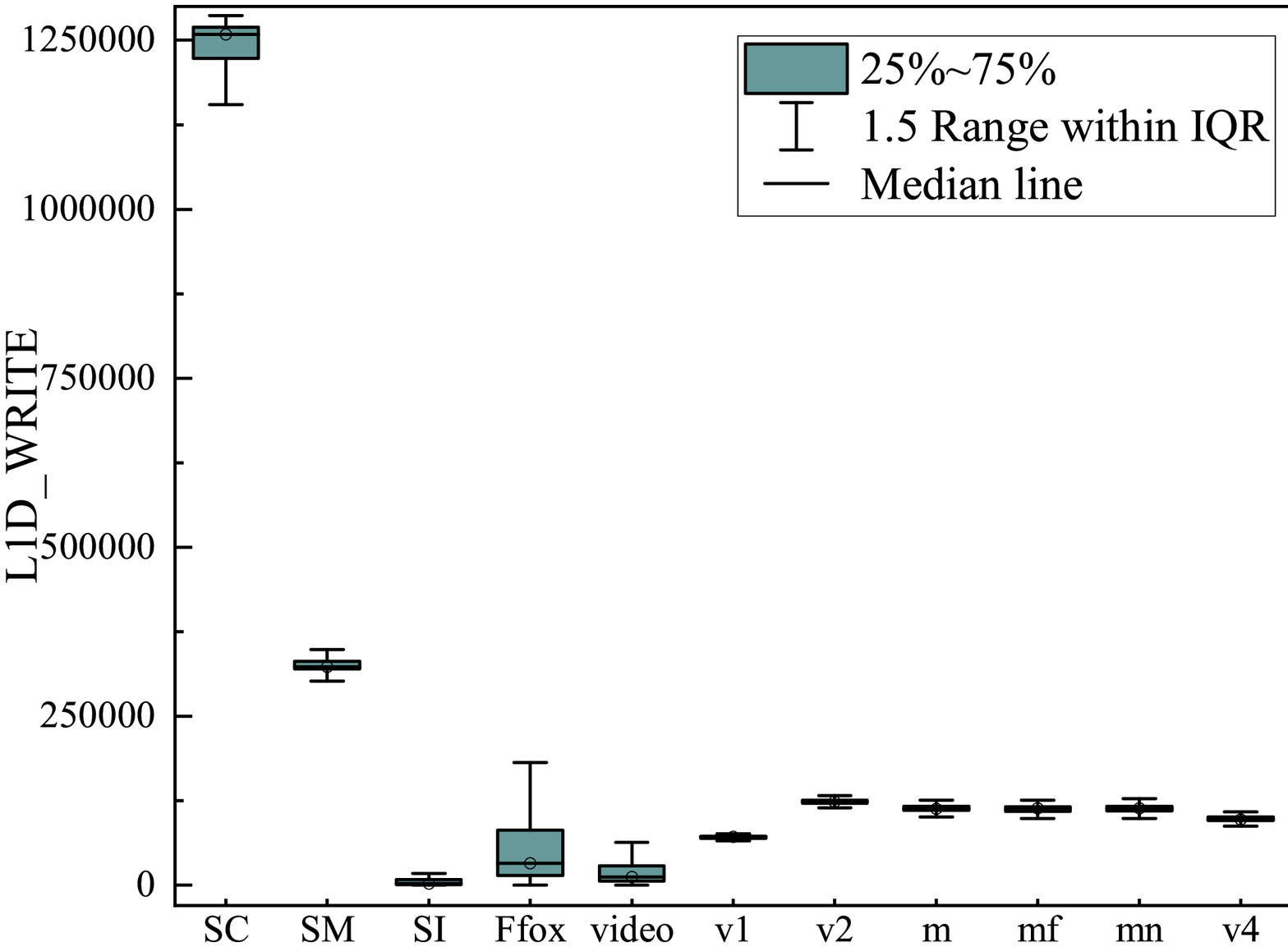}
\centerline{$L1D\_WRITE$}
\end{minipage}
\vfill
\begin{minipage}[t]{0.5\linewidth}
\centering
\includegraphics[width=3.5in]{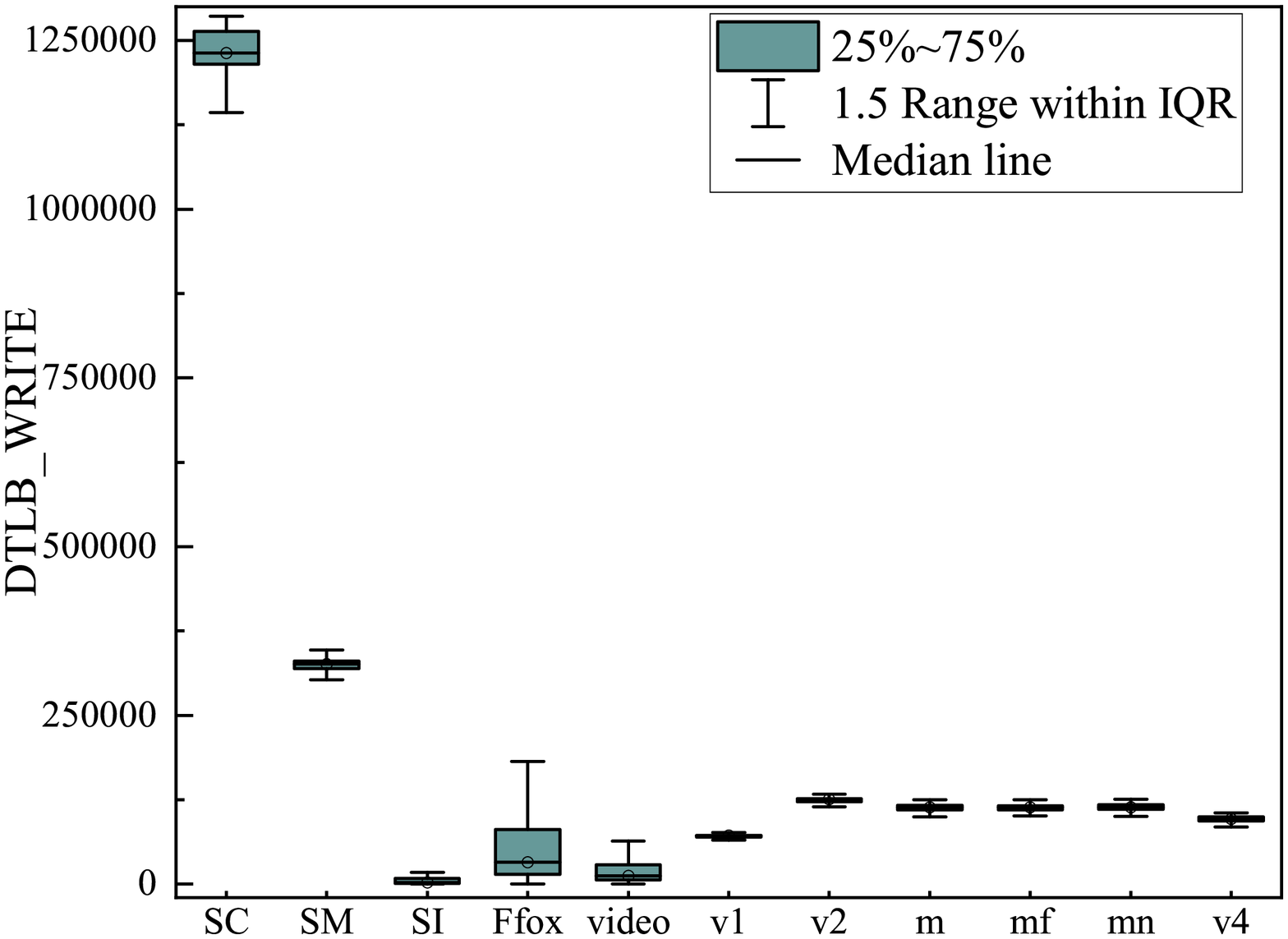}
\centerline{$DTLB\_WRITE$}
\end{minipage}
\begin{minipage}[t]{0.5\linewidth}
\centering
\includegraphics[width=3.5in]{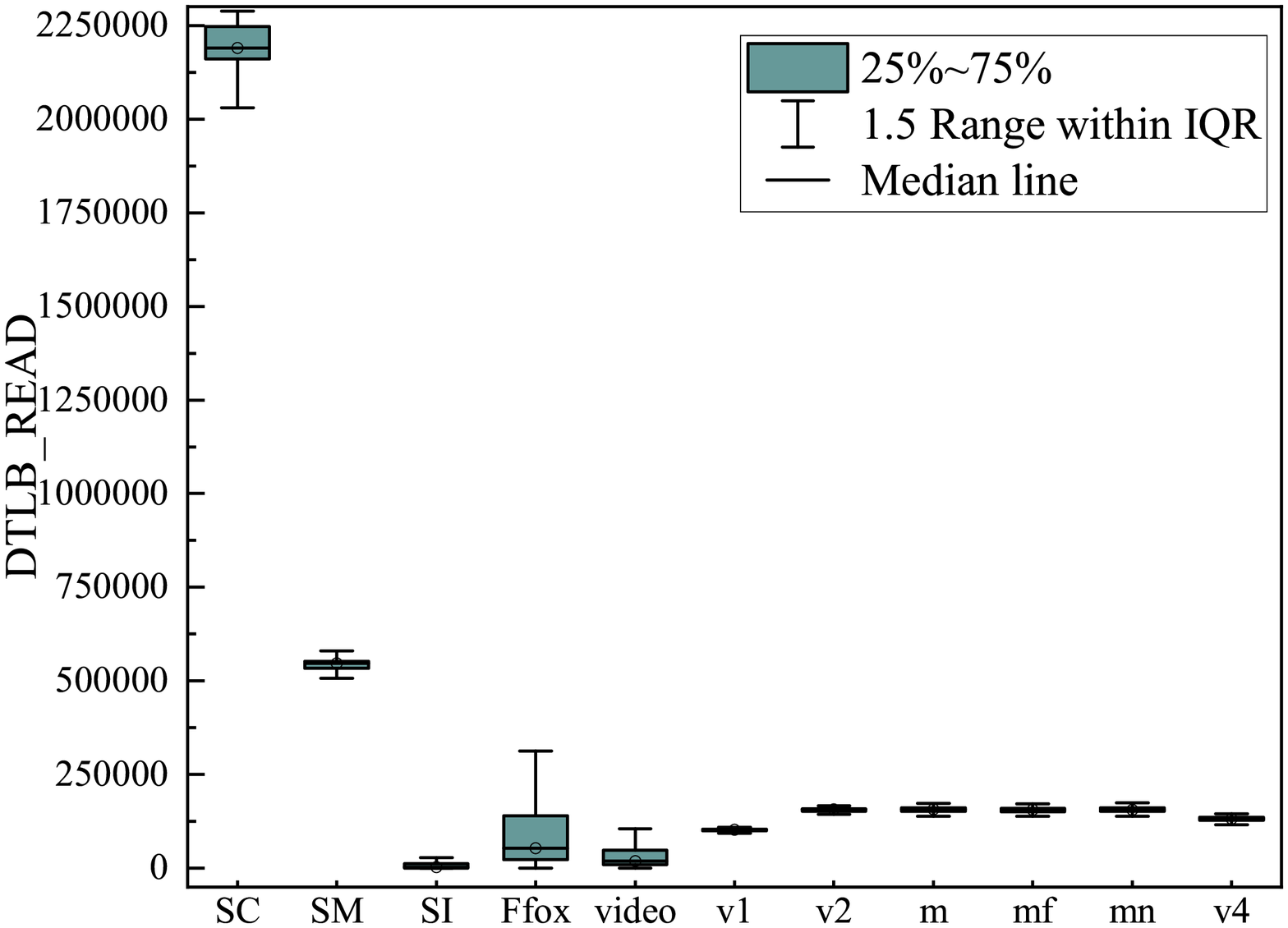}
\centerline{$DTLB\_READ$}
\end{minipage}
\caption{Sample distribution under different scenarios for four features}
\end{figure*}

This paper is the first unified attack detection model built for multiple attacks in contrast to similar works. In this paper, we use four hardware performance events as a feature set and build an attack detection model using Adaboost for multiple variants of the now mainstream Spectre attack and different implementations of meltdown attack, and finally obtain great classification accuracy. The results show that the attack detection mechanism is suitable for discovering the presence of malicious processes in the system, and the detection system is validated in this paper using the validation set with excellent results.

In this paper, to explain the reason for the high accuracy of the obtained attack detection model, the relationship between the features in different scenarios is analyzed, and the four features used for modeling are selected for further study. Figure 5 shows the sample distribution of the same features collected for different scenarios.

By analyzing the sample distribution of different scenarios under the same features, it can be found that there are significant differences between benign and attack programs, which opens up the possibility of attack detection. Moreover, for the attack process, i.e., between V1, V2, Meltdown, and V4, there are also relatively apparent differences between the different features. Previous work often used Page\_fault as the main feature to discover meltdown attacks. However, this paper finds significant differences in page\_fault features under different attack implementations, except for the traditional meltdown attack, which causes a certain number of page errors, and the recent meltdown\_fast meltdown\_nonull attack implementations, which do not have page errors. However, under the four features selected in this paper, the performance of different meltdown attack implementations is similar, and the distribution is significantly different from other scenarios. Hence, the four features selected in this paper are suitable for attack detection against different meltdown attack implementations.

This paper finally implements an attack detection system for multiple Spectre and Meltdown attacks based on a set of four hardware performance events as features with a sampling interval of 1ms and can discover malicious processes in the system accurately. In this paper, the attack detection system also causes only a slight performance loss since it was found in previous work that the performance loss incurred is shallow when the sampling interval is above 500us. We hope to load the whole system into the FPGA, which can further reduce the loss to the system and also improve the versatility of the attack detection module. Accurate attack detection can also be combined with other performance-losing mitigations, such as lfence operations on potentially malicious processes, rather than all processes using the branch predictor, which can save significant performance loss. In future work, since the benign and attack programs in this paper are limited, the utility of attack detection would be significantly enhanced if an autonomous learning module could be added to the model to change the data sample set dynamically. For example, when new samples of benign programs with different feature representations emerge or new samples of Spectre and Meltdown variant attacks emerge, these new samples are added to the original dataset and re-modeled.

\section*{Acknowledgment}

The preferred spelling of the word ``acknowledgment'' in America is without
an ``e'' after the ``g''. Avoid the stilted expression ``one of us (R. B.
G.) thanks $\ldots$''. Instead, try ``R. B. G. thanks$\ldots$''. Put sponsor
acknowledgments in the unnumbered footnote on the first page.


\end{document}